\documentclass[aps,prb,twocolumn,superscriptaddress,showpacs]{revtex4}
\usepackage{amsmath}
\usepackage{bm}
\usepackage{graphicx}
\begin{document}
\title{Conductance and persistent current of a quantum ring coupled to a quantum
wire under external fields}
\author{P. A. Orellana}
\affiliation{Departamento de F\'{\i}sica, Universidad Cat\'{o}lica
del Norte, Casilla 1280, Antofagasta, Chile}
\author{M. L. Ladr\'on de Guevara}
\affiliation{Departamento de F\'{\i}sica, Universidad Cat\'{o}lica
del Norte, Casilla 1280, Antofagasta, Chile}
\author{M. Pacheco}
\affiliation{Departamento de F\'{\i}sica, Universidad T\'ecnica F.
Santa Maria, Casilla 110-V, Valpara\'{\i}so, Chile}
\author{A.Latg\'e}
\affiliation{Instituto de Fisica, Universidade Federal Fluminense,
24210-340, Niter\'oi-RJ, Brazil}

\begin{abstract}
The electronic transport of a noninteracting quantum ring
side-coupled to a quantum wire is studied via a single-band
tunneling tight-binding Hamiltonian. We found that the system
develops an oscillating band with antiresonances and resonances
arising from the hybridization of the quasibound levels of the
ring and the coupling to the quantum wire. The positions of the
antiresonances correspond exactly to the electronic spectrum of
the isolated ring. Moreover, for a uniform quantum ring the
conductance and the persistent current density were found to
exhibit a particular odd-even parity related with the ring-order.
The effects of an in-plane electric field was also studied. This
field shifts the electronic spectrum and damps the amplitude of
the persistent current density. These features may be used to
control externally the energy spectra and the amplitude of the
persistent current.
\end{abstract}


\maketitle
\section{Introduction}

Progress in nanofabrication of quantum devices has allowed one to
study the electron transport through quantum rings in a very
controllable way. Interesting quantum interference phenomena have
been predicted and measured in these mesoscopic systems in
presence of a magnetic flux, such as the Aharonov-Bohm
oscillations in the conductance and persistent
currents.\cite{Chandra,Mailly,Keyser} Also, optical spectroscopy
measurements have allowed a determination of the energy spectra of
closed semiconducting rings.\cite{Lorke} Recently, Fuhrer reported
magnetotransport experiments on closed rings showing the
Aharonov-Bohm effect on the energy spectra.\cite{Fuhrer}

On the other hand, the future miniaturization of electronic
devices have directed attention to the study of discrete
structures, such as arrays of quantum dots and also wires and
rings at the atomic level. \cite{waugh,yanson,kawamura} A recent
experiment reports measurements of the conductance through an
atomic wire placed between two macroscopic contacts, which
exhibits odd-even parity behavior.\cite{smit} This effect was
predicted theoretically,\cite{kim,zeng} and arises of the discrete
nature of the system. \cite{sim} In this article we address a
theoretical study of the transport properties of a quantum ring
side-coupled to a perfect quantum wire in presence of electric and
magnetic fields. The ring may be thought as a chain of quantum
dots or atoms.

The problem of a mesoscopic ring coupled to a reservoir was
discussed theoretically  by B\"uttiker,\cite{Buttiker} in which
the reservoir acts as a source of electrons and an inelastic
scatterer. Takai and Otha considered the case where a magnetic
flux and an electrostatic potential were applied
simultaneously.\cite{Takai} The occurrence of persistent currents
along a normal metal loop connected to two electron reservoirs was
also discussed.\cite{Jayannavar} Moreover the serial of ring
attached to a wire was studied. \cite{gu}All these works are based
on the solutions of the one-electron Schr\"{o}dinger equation for
the ring system, and other systems involving rings have been
studied within a tight-binding model.\cite{damato,liu,Cheung,chen}
This formalism allows a detailed analysis of the variation of the
conductance and the persistent current with the size of the ring.

In contrast to the quantum ring with two contacts, the
transmission through the side-coupled quantum ring consists of the
interference between a ballistic channel (the wire) and the
resonant channels from the quantum ring. Working in the
tight-binding formalism, we show that this system develops an
oscillating band with resonances (perfect transmission) and
antiresonances (perfect reflection). In addition, an odd-even
parity of the number of sites of the ring was found. Namely,
pinning the Fermi energy at the site energy of the quantum ring,
if this number is even perfect transmission takes place and the
persistent current density vanishes for any value of the magnetic
flux. If the number is odd the conductance and the persistent
current density oscillate with the magnetic flux. The effects of
an in-plane electric field applied to the ring on the transport
along the wire waveguide were also investigated. It is shown that
the electric field modulates the position of the resonances and
antiresonances of the linear conductance, and also the period,
amplitude, and phase of the persistent current as a function of
the magnetic-flux oscillation.

\section{Model}
\begin{figure}[h]
\centerline{\includegraphics[width=7cm,angle=0]{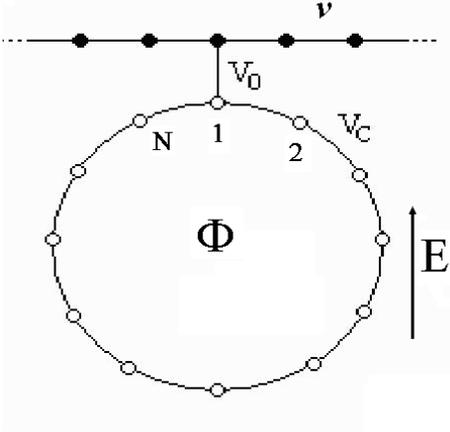}}
\caption{Schematic view of the quantum ring attached to quantum
wire.} \label{fig1}
\end{figure}
The system under consideration is a quantum ring of $N$ atomic
sites connected by tunnel coupling to a quantum wire waveguide, as
depicted schematically in Fig.~\ref{fig1}. A magnetic flux is
assumed to thread the ring and  an electric field applied
perpendicular to the wire is also considered. The full system is
modeled by a single-band tight-binding Hamiltonian within a
noninteracting picture, that can be written as
\begin{equation}
H=H_{W}+H_{R}+H_{WR},
\end{equation}
with
\begin{eqnarray}
H_{W}&=&-v\sum_{<i\neq j>}\,(c_{i}^{\dagger }c_{j}+c_{i}c_{j}^{\dagger }),
\nonumber \\
H_{R}&=&\sum_{l=1}^{N}\varepsilon _{l}d_{l}^{\dagger
}d_{l}-V_{c}(e^{i2\pi \varphi }\,d_{1}^{\dagger }d_{N}+\mbox{h. c.})  \nonumber \\
&+&\sum_{l=1}^{N-1}(V_{c}\,d_{l}^{\dagger }d_{l+1}+\mbox{h. c.}),  \nonumber \\
H_{WR}&=&-V_{0}(d_{1}^{\dagger }c_{0}+\mbox{h. c.}),
\end{eqnarray}

\noindent where $c_{i}^{\dagger}$ and $d_{l}^{\dagger}$ are the
creation operators of an electron at site $i$ and $l$ in the wire
and the ring, respectively, $v$($V_c$) is the corresponding
hopping energy in the wire (ring) and $V_{0}$ is the ring-wire
tunneling coupling. The site energies of the wire and the ring are
set at zero and $\varepsilon _{l}$ respectively. The magnetic flux
is measured in terms of the elemental quantum flux $\Phi
_{o}=hc/e$ as $\varphi =\Phi/\Phi_{0}$. We adopt the singular
gauge for the vector potential associated with the magnetic field,
in which all the effects of the field are included explicitly in
the hopping energy between the first ($1$) and last ($N$) sites of
the ring.\cite{singular}

The eigenstates of the wire Hamiltonian ($H_W$) may be written as
\begin{equation}
\left| k\right\rangle =\sum_{j=-\infty }^{\infty }e^{ikdj}\left|
j\right\rangle ,
\end{equation}
where $d$ is the atomic spacing and $\left|j\right\rangle$ denotes
a Wannier state localized at site $j$. The dispersion relation
associated with these Bloch states reads $\varepsilon(k) = -2v\cos
(kd)$, where the wave vector $k$ is defined within the
corresponding first Brillouin zone $[-\pi /d,\pi /d]$.

The stationary states of the complete Hamiltonian (Eq. (1)) may be written
as
\begin{equation}
\left| \psi _{k}\right\rangle =\sum_{j=-\infty }^{\infty }a_{j}^{k}\left|
j\right\rangle +\sum_{l=1}^{N}b_{l}^{k}\left| l\right\rangle ,
\end{equation}
where the coefficient $a_{j}^{k}$ ($b_{l}^{k}$) is the probability
amplitude to find the electron in the site $j(l)$ of the quantum
wire (ring) in the state $k$.

Solving the eigenvalue problem for $H$ one obtains the following
linear set of coupled equations
\begin{eqnarray}
\varepsilon a_{j}^{k} &=&-v(a_{j-1}^{k}+a_{j+1}^{k})-V_{0}b_{1}^{k}\delta
_{j0},  \nonumber \\
\varepsilon b_{1}^{k} &=&\varepsilon _{1}b_{1}^{k}-V_{c}e^{i2\pi \varphi
}b_{N}^{k}-V_{c}b_{2}^{k}-V_{0}a_{0}^{k},  \nonumber \\
 \varepsilon b_{l}^{k} &=&\varepsilon
_{l}b_{l}^{k}-V_{c}b_{l-1}^{k}-V_{c}b_{l+1}^{k}\quad \mbox{for $l=2,...,N-1$},  \nonumber \\
\varepsilon b_{N}^{k} &=&\varepsilon
_{N}b_{N}^{k}-V_{c}b_{N-1}^{k}-V_{c}e^{-i2\pi\varphi }b_{1}^{k}.
\label{eq5}
\end{eqnarray}
The relationship between the probability amplitudes at the
junction is then given by
\begin{equation}
b_{1}^{k}=\frac{-D_{2,N
}(\varepsilon)}{\widetilde{D}_{N}(\varepsilon)}V_{0}a_{0}^{k},
\label{eq6}
\end{equation}
where $\widetilde{D}_{N}(\varepsilon)$ $=\det(\varepsilon
I-H_{R})$, and $D_{n,m}(\varepsilon)$ is given by
\begin{equation}
D_{n,m}(\varepsilon)=\mbox{det}\left[
\begin{array}{lllll}
\varepsilon -\varepsilon _{n} & V_{c} & 0 & ... & 0 \\
V_{c} & \varepsilon -\varepsilon _{n+1} & V_{c} & ... & 0 \\
0 & V_{c} & ... & ... & ... \\
... & ... & ... & \varepsilon -\varepsilon _{m-1} & V_{c} \\
0 & ... & 0 & V_{c} & \varepsilon -\varepsilon _{m}
\end{array}
\right].
\end{equation}

Following standard methods of quantum waveguide transport, one may
calculate the transmission coefficient and obtain the probability
amplitudes $a_{j}^{k}$ via an iterative procedure.\cite{pedro1} As
usual, electrons are described by a plane wave incident from the
far left with unit amplitude and a reflection amplitude $r$, and
at the far right by a transmission amplitude $t$. For a given
transmission amplitude, the associated incident and reflection
amplitudes may be determined by matching the iterated function to
the proper plane wave at the far left. For $a_0^k$ one gets
%\begin{subequations}
\begin{eqnarray}
a_0^k=t&=&\frac{2iv\sin (kd)}{2iv\sin (kd)-V_{0}^{2}D_{2,N}(\varepsilon)/\widetilde{D}_{N}(\varepsilon)}\\
&=&\frac{%
\widetilde{D}_{N}(\varepsilon)}{\widetilde{D}_{N}(\varepsilon)+i\Gamma
D_{2,N}(\varepsilon)}, \label{eq8}
\end{eqnarray}
%\end{subequations}
\smallskip
\noindent where $\Gamma (k)\equiv \Gamma_0 /\sin (kd)$  , with
$\Gamma_0 = V_{0}^{2}/2v$. The transmission probability is given
by $T=\left| t\right| ^{2}$.

The linear conductance at the Fermi level is calculated via the
one-channel Landauer formula at zero temperature. Actually, the
conductance is the experimentally accessible quantity related to
the transmission probability $T$,
\begin{equation}
G(\varepsilon)=\frac{2e^{2}}{h}\,T(\varepsilon)=\frac{2e^{2}}{h}\,\frac{\left| \widetilde{D}%
_{N}(\varepsilon)\right| ^{2}}{\left|
\widetilde{D}_{N}(\varepsilon)\right| ^{2}+\Gamma ^{2}\left|
D_{2,N}(\varepsilon)\right| ^{2}}. \label{Eq8}
\end{equation}
One observes that $G(\varepsilon)$ vanishes when $%
\widetilde{D}_{N}(\varepsilon)$ is zero, and is equal to $2e^2/h$
for null values of $D_{2,N}(\varepsilon)$. One should notice that
the zeros of $\widetilde{D}_{N}(\varepsilon)$ correspond to the
energy spectrum of the isolate ring. Thus, the energy spectrum of
a particular ring configuration may be obtained by measuring the
zeros of the conductance.

In the case of a magnetic flux threading the ring, one knows that
a persistent current is generated through the circular system. The
persistent current density $J$ along the ring, in the energy
interval $d\varepsilon $ around $\varepsilon$, is obtained from
\cite{pedro2}
\begin{equation}
J=\frac{2eV_c}{\hbar}\,Im(b_{l+1}^{*}b_{l}). \label{pc}
\end{equation}
As this quantity does not depend on the site $l$, one may evaluate
it between any pair of coupled sites. For simplicity, we choose
$l=1$ and $2$. It follows from Eqs. (\ref{eq5}) that for a ring of
$N$ sites, $b_2^k$ is given by
\begin{equation}
b_{2}^{k}=V_{0}V_{c}\frac{(D_{3,N}(\varepsilon)+\cos (N\pi
)e^{-i2\pi \varphi
}V_{c}^{N-2})}{\widetilde{D}_{N}(\varepsilon)}a_{0}^{k},
\end{equation}
which together with Eqs. (\ref{eq6}), (\ref{eq8}) and (\ref{pc})
gives
\begin{equation}
J=-\frac{2ev}{\hbar}\,\sin (2\pi \varphi )\frac{\cos (N\pi
)\Gamma_0 V_{c}^{N}D_{2,N}(\varepsilon)}{\left|
\widetilde{D}_{N}(\varepsilon)\right| ^{2}+\Gamma ^{2}\left|
D_{2,N}(\varepsilon)\right| ^{2}}. \label{Eq10}
\end{equation}
It is worth noting that when $D_{2,N}=0$ and $\widetilde{D}_N\neq
0$, the transmission is perfect ($G=2e^2/h$) and the persistent
current density vanishes. On the other hand, when
$\widetilde{D}_{N}=0$, the linear conductance vanishes and the
persistent current density oscillates regularly with the magnetic
flux with period $\Phi_0$.

\section{Zero electric field}
Let us introduce the dimensionless conductance $g=G/(2e^2/h)$ and
persistent current density $j=J/(2ev/\hbar)$. For zero electric
field, we assume here the particular case of a uniform ring where
the ring site energies are $\varepsilon _{l}=\varepsilon_o=0$ (for
$l=1,...,N$). The dimensionless conductance $g(\varepsilon)$ can
be written in a compact form, which depends explicitly on the size
of the ring,
\begin{figure}[t]
\centerline{\includegraphics[width=7cm,angle=-90,clip=]{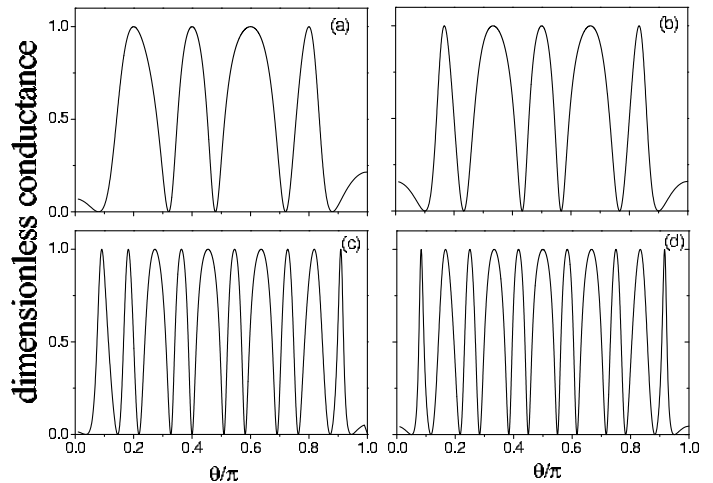}}
\caption{Dimensionless conductance as a function of $\theta$
($\theta=\arccos(\varepsilon/2V_c)$), for rings composed of (a) 5,
(b) 6, (c) 11, and (d) 12 atomic sites) and for a magnetic flux
equal to 0.3.} \label{fig2}
\end{figure}
\begin{figure}[t]
\centerline{\includegraphics[width=60mm,clip=]{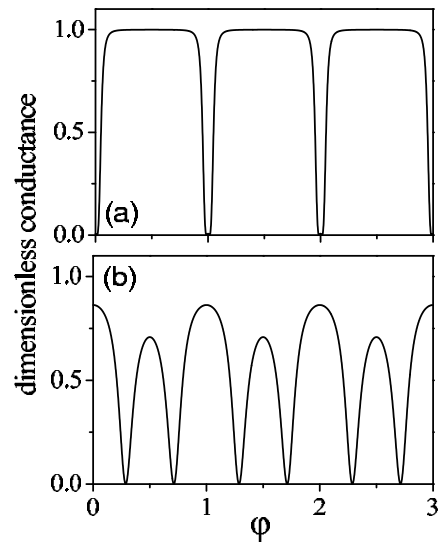}}
\caption{Dimensionless conductance as a function of the magnetic
flux $\varphi=\Phi/\Phi_o$ for rings composed of (a) 11, and (b)
12 atomic sites.} \label{fig3}
\end{figure}
\begin{equation}
g(\varepsilon)= \frac{1}{1+h(\varepsilon)},
\end{equation}
where
\begin{equation}
h(\varepsilon)=\frac{(\prod\limits_{i=1}^{N-1}[\varepsilon
-2V_{c}\cos (\pi
i/N)])^{2}\Gamma^{2}}{(\prod\limits_{i=0}^{N-1}\{\varepsilon
-2V_{c}\cos [2\pi (i+\varphi )/N]\})^{2}}
\end{equation}
\noindent The occurrence of resonances ($g(\varepsilon)=1$) are
then expected at $\varepsilon =2V_{c}\cos(\pi i/N)$
($i=1,...,N-1$) and are independent of the magnetic flux, whereas
antiresonances (that is, $g(\varepsilon)=0$) take place at
energies $\varepsilon =2V_{c}\cos[2\pi (i+\varphi )/N]$
($i=0,...,N-1$). Introducing the energy parameter defined by
$\theta =\arccos(\frac{1}{2}\varepsilon /V_{c})$, $h(\varepsilon
)$ may be written (see details in the Appendix) as
\begin{equation}
h(\varepsilon )=  \frac{(\Gamma/2V_{c})^{2}\sin (N\theta )^{2}}
{\sin(\theta)^2[\cos (N\theta )-\cos (N\pi )\cos (2\pi \varphi
)]^{2}}. \label{eq11}
\end{equation}
\begin{figure}[t,h,b]
\centerline{\includegraphics[width=6cm,angle=-90,clip=]{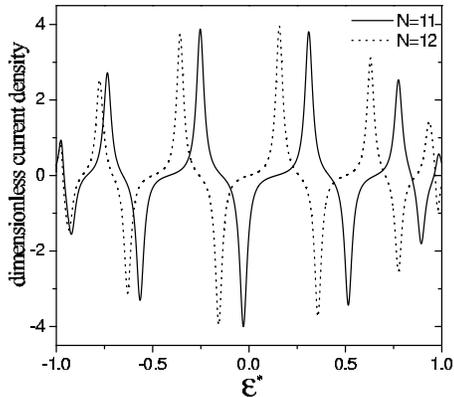}}
\caption{Dimensionless current density as a function of the Fermi
energy for rings with  N$=11$ (solid curve) and N$=12$ (dot line)
atomic sites, and for a magnetic flux equal to $0.3$.}
\label{fig4}
\end{figure}

Results for the conductance as a function of $\theta$ are shown in
Figure~\ref{fig2}  for rings of different sizes and considering a
magnetic flux $\varphi=0.3$ and $\Gamma_0=V_{c}$. The conductance
clearly exhibits an oscillating pattern of resonances and
antiresonances for particular energies which depend on the number
of sites in the ring. As mentioned above, the corresponding
antiresonant energies give us the energy spectrum of the ring. The
quantum wire conductance dependence on the magnetic flux is
explicitly shown in Figure~\ref{fig3} for two values of $N$
($N=11$ and 12), considering a Fermi energy equal to $0.3V_c$. The
oscillatory period of a quantum of flux is found, independent of
the ring order, as expected from the analytical expression for the
conductance [Eq. (\ref{eq11})]. This behavior is compatible with
the results found by Shi and Gu for the case of one ring
side-attached to leads.\cite{gu} Let us now calculate the
persistent current density. For a regular ring this reduces to
\begin{widetext}
\begin{equation} j(\varepsilon)=\frac{(\Gamma_0 /2V_{c})\cos (N\pi)\sin
(N\theta) \sin(\theta)\sin (2\pi \varphi )}{\sin^2(\theta)[\cos
(N\theta)-\cos (N\pi )\cos (2\pi \varphi ) ]^{2}+(\Gamma
/2V_{c})^{2}\sin (N\theta )^{2}}.
\end{equation}
\end{widetext}
One clearly notices that $j$ is an oscillating function of the
Fermi energy ($\theta$) and of the magnetic flux ($\varphi$).
Moreover, in the limit $\Gamma\rightarrow 0$ the current density
exhibits a delta function behavior in the energies of the isolated
ring (this limit correspond to the disconnected ring).
\begin{figure}[h]
\centerline{\includegraphics[width=60mm,angle=-90,clip=]{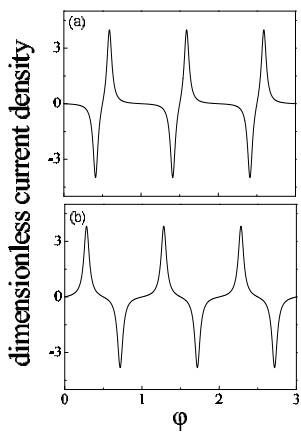}}
\caption{Dimensionless current density as a function of the
magnetic flux for a Fermi energy equal to 0.1$V_c$ and for (a)
$N=11$ and (b) $N=12$.} \label{fig5}
\end{figure}

It is straightforward to show from Eq. (\ref{eq11}) that when the
Fermi energy is pinned at zero ($\theta =\pi /2$), the
dimensionless conductance $g$ reduces to
\begin{eqnarray}
g &=&\frac{\cos (2\pi \varphi )^{2}}{\cos (2\pi \varphi )^{2}+
\left( \Gamma /2V_{c}\right) ^{2}},\quad N \text{ odd}  \nonumber \\
g\, &=&\,1,\quad N \text{ even},
\end{eqnarray} and the
corresponding persistent current density is
\begin{eqnarray}
j &=& \frac{(\Gamma_0/2V_c) \sin (2\pi \varphi )}{\cos (2\pi
\varphi )^{2}+\left( \Gamma /2V_{c}\right) ^{2}},\quad N \text{
odd}\nonumber
\\
j &=&\,0,\quad N \text{ even}.
\end{eqnarray}
\begin{figure}[h]
\centerline{\includegraphics[width=60mm,clip=]{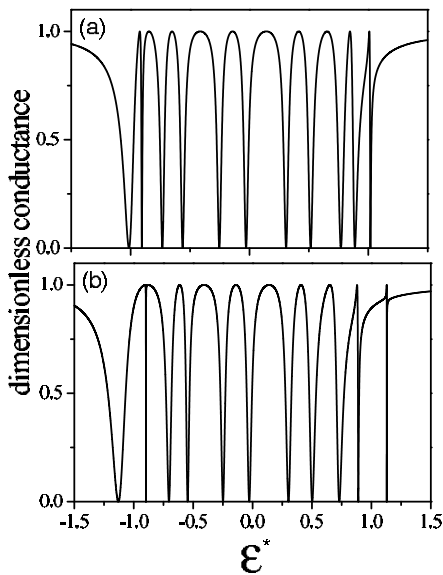}}
\caption{Dimensionless conductance as a function of the Fermi
energy, for  a  ring with N$=11$, a magnetic flux $\varphi =0.3$,
and for distinct electric field intensities, (a) $E^*=0.05$, and
(b) $E^*=0.15$} \label{fig6}
\end{figure}
Notice that for a uniform ring of even $N$, perfect transmission
takes place ($g=1$) and the persistent current density vanishes
($j=0$) for any value of the magnetic flux. The transmission is
perfect in this case because for this energy the electron does not
enter the ring (in fact, it can be shown that its phase remains
unaltered), and therefore the magnetic flux does not play any role
in the conductance. This also explains that the persistent current
density is zero. For $N$ odd, the transmission and the persistent
current density oscillate regularly with the magnetic flux.

The persistent current density as a function of the dimensionless
energy $\varepsilon^*$ ($\varepsilon^*=\varepsilon/2V_c$) is
depicted in Figure 4 for two ring configurations ($N=11$ and
$12$), for a magnetic flux $\varphi=0.3$. The expected oscillatory
behavior with the energy is clearly evidenced, as well as the
dependence on the magnetic flux, as shown in Figure~\ref{fig5} for
the fixed energy $\varepsilon^* =0.1$, and for both an odd and an
even ring number configuration.
\begin{figure}[!]
\centerline{\includegraphics[width=60mm,clip=]{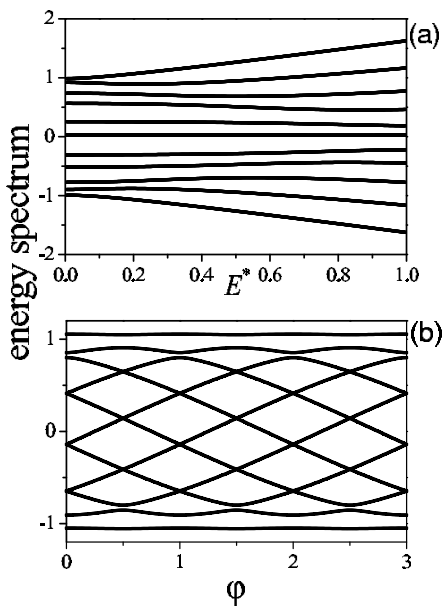}}
\caption{Electronic spectrum of an 11 sites ring as a function of
a) the electric field strength, for a magnetic flux  of
$\varphi=0.3$, b) the magnetic flux, for an electric field
strength of $E^*=0.1$. The energy is in units of $2V_c$.}
\label{fig7}
\end{figure}

\section{Electric field effects}

Within the tight-binding approximation, and assuming an in-plane
ring electric field perpendicular to the wire, the dependence of
the site energy on the field may be expressed as $\varepsilon
_{l}=(eEdN/2\pi)\cos(2\pi (l-1)/N)=(eV_c)(E^*N/2\pi)\cos(2\pi
(l-1)/N)$, where we define the dimensionless electric field
strength $E^*=Ed/V_c$. The determinants
$\widetilde{D}_{N}(\varepsilon)$ and $D_{1,N}(\varepsilon)$ are
now calculated iteratively (see details in the Appendix).
Figure~\ref{fig6} shows the dimensionless conductance as a
function of the Fermi energy, for a ring with $N=11$ sites,
magnetic flux $\varphi =0.3$, and two  electric field values. The
main effects of the electric field is to shift and squeeze the
resonances and antiresonances of the linear conductance.

The energy spectrum of an isolated ring ($N=11$) in the presence
of both magnetic and electric field is also analyzed. The
dependence of the spectrum on the electric field energy is
displayed in Figure (\ref{fig7}-a), for $\varphi = 0.3$, whereas
Figure (\ref{fig7}-b) shows the explicit dependence on the
magnetic flux for an electric field $E^*=0.1$. One of the main
effects of the electric field is the suppression of the
Aharonov-Bohm oscillations in the edges of the energy spectra,
being as the lowest and highest-lying energy levels are almost
independent on the magnetic flux. For energies in the center of
the energy spectrum new oscillations with the quantum-flux period
are developed. Quite similar results were found before for the
case of a finite-width semiconducting quantum ring.\cite{pupi}

\begin{figure}[h]
\centerline{\includegraphics[width=60mm,angle=-90,clip=]{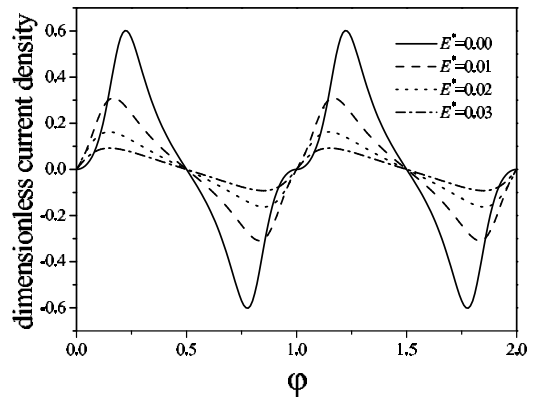}}
\caption{Dimensionless current density in $N=11$ sites ring as a
function of magnetic flux for different values of electric field,
$E^*=0.0$ (solid line), $E^*=0.1$ (dash line), $E^*=0.2$ (dot
line) and $E=0.3^*$ (dash-dot line)} \label{fig8}
\end{figure}

\begin{figure}[h]
\centerline{\includegraphics[width=60mm,angle=-90,clip=]{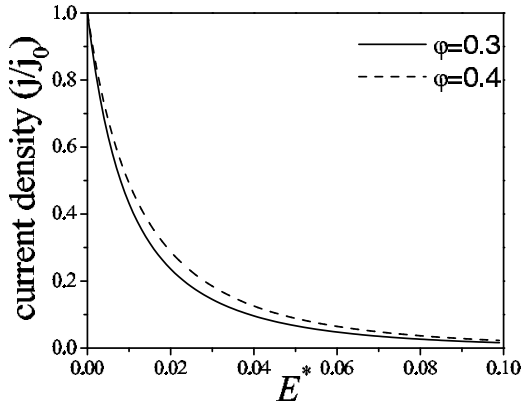}}
\caption{Dimensionless current density in $N=11$ sites ring as a
function of electric field strength for fixed magnetic flux
$\varphi=0.3$ (solid line) and $\varphi=0.4$ (dash line).}
\label{fig9}
\end{figure}

As expected, the current density is also affected by the electric
field. Figure~\ref{fig8} shows the dimensionless current density
as a function of the magnetic flux for different values of the
electric field. As we can appreciate, the persistent current
density decreases with the electric field strength. It is shown
more clearly in Figure~\ref{fig9}. This figure shows the
normalized current density $j/j_0$ associated with the lowest
level of the energy spectrum as a function of the electric field
strength for for a fixed magnetic flux ($j_0$ current density at
zero magnetic flux). The current density decays exponentially with
the electric field strength. This way, the electric field can be
used to control the persistent current in quantum rings.

\section{Summary}
The conductance and the persistent current density, at zero
temperature, of a side ring attached to a quantum wire was
investigated. We found that the system develops an oscillating
band with antiresonances and resonances arising from the
hybridization of the quasibound levels of the ring and the
coupling to the quantum wire. The positions of the antiresonances
correspond exactly to the electronic spectrum of the isolated
ring. For a uniform ring we found that the system shows odd-even
parity in the conductance and also in the persistent current
density. Namely, when the Fermi energy is pinned at zero, if the
number of sites in the ring is even, perfect transmission takes
place and the persistent current density vanishes for any value of
the magnetic field. When this number is odd, the transmission and
current density oscillate periodically with the magnetic flux. The
effects of an in-plane electric field were also studied. We found
that this shifts the electronic spectrum and damps the amplitude
of the persistent current density. These features may be used to
control externally the energy spectra and the amplitude of the
persistent current.

\section*{ACKNOWLEDGMENTS}

P.\ A.\ O.\ and M.\ P.\ would like to thank financial support from
Milenio ICM P99-135-F, FONDECYT under grant 1020269, 1010429, and
7010429 and Red IX.E "Nanoestructuras para la Micro y
Optoelectronica Sub-Programa IX "Microelectronica" Programa CYTED.
A. L. was partially supported by the Brazilian Agencies CNPq and
FAPERJ.

\section*{APPENDIX}

%\subsection{Appendix}

We present here the properties of the determinant
$D_{n,m}(\varepsilon)$ which is defined by Eq. (\ref{eq5}). It is
easy to show that the following recurrence relation is valid
\begin{eqnarray}
D_{n,m}=(\varepsilon -\varepsilon
_{n})D_{n+1,m}-V_{c}^{2}D_{n+2,m},\\\nonumber
 n = 1,2...,m,m=3,4,...
\end{eqnarray}
with $D_{m,m}=\varepsilon -\varepsilon _{m}$ and $D_{m-1,m}$
$=(\varepsilon -\varepsilon _{m-1})(\varepsilon -\varepsilon
_{m})-V_{c}^{2}$. Mainly for a uniform case $\varepsilon _{m}=0$
for all $m^{^{\prime }}s$, the expression for $D_{n,m}\equiv
D_{n}$ can be found in a explicit form. In fact,
$D_{n}(\varepsilon )$ is related with the type-II Chebyshev
polynomial,

\begin{equation}
D_{n}(\varepsilon )=V_{c}^{n}U_{n}(\frac{1}{2}\varepsilon /V_{c}).
\end{equation}

Then it is straightforward to show that
\begin{equation}
D_{n}=V_{c}^{n}\frac{\sin ((n+1)\theta )}{\sin \theta }=\prod%
\limits_{i=1}^{n}(\varepsilon -2V_{c}\cos (\pi i/(n+1))).
\end{equation}

The determinant $\widetilde{D}_{N}$ can be written in function of
the determinant $D_{n,m}$ as

\begin{eqnarray}
\nonumber \widetilde{D}_{N}&=&(\varepsilon -\varepsilon
_{1})D_{2,N}-V_{c}^{2}(D_{3,N}+D_{2,N-1})\\
 &-&2V_{c}^{N}\cos
(N\pi )\cos (2\pi \varphi ).
\end{eqnarray}

In the particular case when $\varepsilon _{m}=0$ for all $m^{^{\prime }}s$,
the expression for $\widetilde{D}_{N}$ can be written in terms of the
eigenvalues of $H_{N}$ $(\varepsilon _{i}=2V_{c}\cos (2\pi (i+\varphi )/N)).$

\begin{equation}
\widetilde{D}_{N}=\prod\limits_{i=1}^{N}(\varepsilon -2V_{c}\cos (2\pi
(i+\varphi )/N)).
\end{equation}

Additionally, from Eq.(21) we can write the determinant
$\widetilde{D}_{N}$ as a function of $\theta$,
%\begin{widetext}
\begin{eqnarray}
\nonumber \widetilde{D}_{N} &=&\varepsilon
D_{N-1}-2V_{c}^{2}D_{N-2} - V_{c}^{N}\cos (N\pi) \cos (2\pi
\varphi)\\
 &=& 2V_{c}^{N}[\cos (N\theta )- \cos(N\pi )\cos(2 \pi
\varphi )].
\end{eqnarray}
%\end{widetext}


\begin{thebibliography}{99}


\bibitem{Chandra}  V. Chandrasekhar, R.A. Webb, M.J. Brady, M.B. Ketchen, W.J.
Gallagher and A. Kleinsasser, Phys. Rev. Lett. \textbf{67} 3578
(1991).

\bibitem{Mailly}  D. Mailly, C. Chapelier and A. Benoit, Phys. Rev Lett.
\textbf{70} 2020 (1993).

\bibitem{Keyser}  U. F. Keyser, C. F\"{u}hner, S. Borck and R. J. Haug,
Semm. Sc. and Tech. {\bf 17}, L22 (2002).

\bibitem{Lorke}  A. Lorke, R. J. Luyken, A. O. Govorov, J\"{u}rg P.
Kotthaus, J. M. Garc\'{i}a, P. M. Petroff, Phys. Rev. Lett.
\textbf{84} 2223 (2000).

\bibitem{Fuhrer}  A. Fuhrer, S. L\"{u}scher, T. Ihn, T. Heinzel, K. Ensslin,
W. Wegscheiner, M. Bichler, Nature \textbf{413} 385 (2001).

\bibitem{yanson} A.I. Yanson, G. Rubio-Bollinger, H. E. van den
Brom, N. Agra\"\it, and J.M. van Ruitenbeek, Nature
(London),\textbf{395}, 780 (1998).
\bibitem{waugh} F.R. Waugh, M.J. Berry, C.H. Crouch, C. Livermore,
D.J. Mar, and R.M. Westervelt, K.L. Campman and A.C. Gossard,
Phys. Rev. B. \textbf{53} 1413 (1996).
\bibitem{kawamura}Midori Kawamura, Neelima Paul, Vasily Cherepanov, and Bert Voigtlan\"ander,
Phys. Rev. Lett. \textbf{91}, 096102 (2003).
\bibitem{smit} R.H.M. Smit, C. Untiedt, G. Rubio-Bollinger, R.C.
Segers, and J.M. van Ruitenbeek, Phys. Rev. Lett. \textbf{91},
076805 (2003).
\bibitem{zeng} Z.Y. Zeng  and F.Claro, Phys. Rev. B \textbf{65},
193405 (2002).
\bibitem{kim} T.S. Kim and S. Hershfield, Phys. Rev. B.
\textbf{65}, 214526 (2002).
\bibitem{sim} H.-S. Sim, H.-W. Lee, and K.J. Chang, Phys. Rev. Lett. \textbf{87}, 096803 (2001).

\bibitem{Buttiker}  M. B\"{u}ttiker, Phys. Rev. B \textbf{32} 1846 (1985).

\bibitem{Takai} D. Takai and K. Ohta, J. Phys. Cond. Matt. \textbf{6}, 5485 (1994); {\it ibid} Phys. Rev. B {\bf 48}, 14318 (1993).

\bibitem{Jayannavar} A. M. Jayannavar and P. Singha Deo, Phys. Rev. B \textbf{49}, 13685 (1994).

\bibitem{gu} Ji-Rong Shi and Ben-Yuan Gu, Phys. Rev. B \textbf{55}, 4703 (1997).

\bibitem{damato} Jorge L. D'Amato, Horario M. Pastawski, and Juan
F. Weisz, Phys. Rev. B \textbf{39}, 3554 (1989).
\bibitem{liu} Youyan Liu and P.M. Hui, Phys. Rev. B
\textbf{57},12994 (1998).
\bibitem{Cheung} Ho-Fai Cheung and Eberhard K. Riedel, Phys. Rev. B \textbf{40}, 9498 (1989).
\bibitem{chen} Yan Chen, Shi-Jie Xiong, S. N. Evangelou, Phys.
Rev. B \textbf{56}, 4778 (1997).


\bibitem{singular} J. P. Carini, K. A. Muttalib, and S. R. Nagel,
Phys. Rev. Lett. \textbf{53}, 102 (1984).

\bibitem{pedro1} P. A. Orellana, F. Dominguez-Adame, I. Gomez, and M. L. Ladr\'on de Guevara,
  Phys. Rev. B \textbf{67}, 085321 (2003).

\bibitem{pedro2} Pedro A. Orellana, G.A. Lara and Enrique V. Anda,
  Phys. Rev. B \textbf{65}, 155317 (2002).

\bibitem{pupi}  Z. Barticevic, G. Fuster and M. Pacheco, Phys. Rev. B
\textbf{65} 193307 (2002).
\end{thebibliography}
\end{document}